# Investigation of Complex Impedance and Modulus Properties of Nd Doped 0.5BiFeO$_3$-0.5PbTiO$_3$ Multiferroic Composites

Ajay Kumar Behera, N.K. Mohanty, S.K. Satpathy, Banarji Behera[*] and P. Nayak

Materials Research Laboratory, School of Physics, Sambalpur University,

Jyoti Vihar, Burla – 768019, Odisha, India.

**Abstract**

0.5BiNd$_x$Fe$_{1-x}$O$_3$-0.5PbTiO$_3$ (x=0.05, 0.10, 0.15, 0.20) composites were successfully synthesized by a solid state reaction technique. At room temperature X-ray diffraction shows tetragonal structure for all concentrations of Nd doped 0.5BiFeO$_3$-0.5PbTiO$_3$ composites. The nature of Nyquist plot confirms the presence of bulk effects only for all compositions of Nd-doped 0.5BiFeO$_3$-0.5PbTiO$_3$ composites. The bulk resistance is found to decreases with the increasing in temperature as well as Nd concentration and exhibits a typical negative temperature coefficient of resistance (NTCR) behavior. Both the complex impedance and modulus studies have suggested the presence of non-Debye type of relaxation in the materials. Conductivity spectra reveal the presence of hopping mechanism in the electrical transport process of the materials. The activation energy of the composite increases with increasing Nd concentration and were found to be 0.28, 0.27, 0.31 and 0.32eV for x=0.05, 0.10, 0.15, 0.20 respectively at 200-275 $^o$C for conduction process.

Keywords: X-ray Diffraction, Composite Materials, Electrical Impedance, Relaxation Phenomena.

**PACS** 61.05.cp, 77.84.Lf, 87.63.Pn, 67.25.du

---

[*] Corresponding author: Tel.: +91 9439223383, E-mail Id: banarjibehera@gmail.com

# 1 Introduction

Recently, there has been an extensive study in the direction of newer materials possessing ferromagnetic as well as the ferroelectric properties because of the richness of physics involved in the system as well as their potential applications in memory and functional devices [1-6]. These materials can be considered as multiferroic with the interaction of electric and magnetoelectric effects, and the effect of mutual influence of the polarization and magnetization [3,4,7]. $BiFeO_3$ (BFO) is one of the natural ferroelectromagnets which exhibit both ferroelectric and ferromagnetic ($T_C$=1103 K and $T_C$=1103 K) and shows a rhombohedrally distorted perovskite crystal structure with space group of R3c [8], which makes it a promising material for application at room temperature. However, its low remanent magnetization and relatively leakage current are the main limitation for possible device application. Hence, the special attention has been laid on synthesis of BFO with some rare earth or other group element as reviewed by Silva et al [9] for improved multiferroic properties. However, there are reports of coexistence of ferroelectricity and magnetism in BFO which is induced by the rare earth substitution in the A and B site of the perovskite structure. Rare earth ions such as $La^{3+,}$ $Sm^{3+}$ and $Nd^{3+}$ have been added at the A-site and $Nb^{5+}$, $Mn^{4+}$, $Ti^{4+}$ or $Cr^{3+}$ at the B-site for enhancing the electrical and magnetic properties of BFO [10-15]. The series of $(BiFeO_3)_x$– $(PbTiO_3)_{1-x}$ solid solution was first reported on by Venevstev *et al.* in 1960 [16]. They observed that the tetragonal phase persists in the series from the $PbTiO_3$ end member up to 70 wt% BFO and remarked upon the large *c/a* ratio present in the tetragonal structure at 60 wt% BFO. Moreover, Wang et al. [17] reported an enhancement of ME properties in the (1-x) $BiFeO_3$-$xPbTiO_3$ solid solutions. Zhu et al. [18] proposed a phase diagram for the (1-x)BFO-xPT. Their results reveal the existence of a morphotropic phase boundary (MPB) in this system, at which rhombohedral (x≤0.20), orthorhombic (0.20≤x≤0.28) and tetragonal (x ≥ 0.31) phases exist with a large tetragonality in the tetragonal phase region. Further the impedance, modulus, relaxation time, ac and dc conductivity properties have not been investigated so far at elevated temperatures. In this work, we report the complex impedance and modulus properties of $0.5BiNd_xFe_{1-x}O_3$-$0.5PbTiO_3$ (x=0.05, 0.10, 0.15, 0.20) multiferroic composites.

## 2 Experimental

Neodymium (Nd) doped solid solutions of 0.5BFO-0.5PT with the formula $0.5BiNd_xFe_{1-x}O_3$-$0.5(PbTiO_3)$ (x=0.05, 0.10, 0.15, 0.20) were prepared by the solid state reaction technique. The high purity precursors: $Bi_2O_3$ and $Nd_2O_3$ (99.99 %, Spectrochem Pvt. Ltd., India), $Fe_2O_3$ (≥ 99%, M/s Loba Chemicals, Pvt. Ltd., India), PbO (≥95%, Qualigens Pvt. Ltd., India) and $TiO_2$ (≥ 99% Merck Specialties Pvt. Ltd., India) were carefully weighed in a suitable stoichiometric proportion. The polycrystalline materials were mixed thoroughly and ground in an agate mortar for 2 h and in methanol for 2 h. The mixed powders were calcined in a high purity alumina crucible at an optimized temperature of 800 °C for 6h in an air atmosphere. The process of grinding and calcinations was repeated several times till the formation of the compound was confirmed. Then calcined powder was mixed with polyvinyl alcohol (PVA) which acts as a binder to reduce the brittleness of the pellet and burnt out during high temperature sintering. The fine homogenous powder was cold pressed into cylindrical pellets of 10 mm diameter and 1-2 mm of thickness at pressure of $4x10^6$ $N/m^2$ using a hydraulic press. These pellets were sintered at 800 °C for 6 h in an air atmosphere. The sintered pellets were polished by fine emery paper to make both the surfaces flat and parallel. To study the electrical properties of the composites, both flat surfaces of the pellets were electroded with air-drying conducting silver paste. After electroding, the pellets were dried at 150 °C for 4 h to remove moisture, if any, and then cooled to room temperature before taking any electrical measurement. The formation of composites were studied by an X-ray diffraction technique at room temperature with a powder diffractometer (Rigaku Miniflex, Japan) using $CuK_α$ radiation ($λ$ = 1.5405 Å) in a wide range of Bragg's angles $2θ(20^0≤θ≤80^0)$ with a scanning rate of $3°/min$. The electrical parameters (impedance, modulus and capacitance) of the composites were measured using an LCR meter (HIOKI, Model-3532) in the frequency range of $10^2$-$10^6$ Hz from 25-450 °C.

## 3 Results and discussion

### 3.1 Structural analysis

Figure 1 shows the XRD patterns of $0.5BiNd_xFe_{1-x}O_3$-$0.5PbTiO_3$ ($BN_xF_{1-x}$-PT) with x = 0.05, 0.10, 0.15, 0.20 at room temperature. The tetragonal crystal structure has been confirmed for all concentrations ranging from x=0.05 - 0.20. The composites show large tetragonality with increase in Nd concentration from 0.05-0.20 as it is expected from pure $BiFeO_3$-$PbTiO_3$ multiferroics composites [19]. It was found good agreement between observed (obs) and calculated (cal) interplanar spacing d ($\sum \Delta d = d_{obs} - d_{cal}$ = minimum). The lattice parameters of the selected unit cells were refined using the least-squares sub-routine of a standard computer program package "POWD" [20]

**3.2 Impedance Properties**

Complex impedance spectroscopy (CIS) [21] is a unique and powerful technique to characterize the electrical behavior of a system. This analysis enables one to resolve the contributions of various processes such as the bulk, grain boundaries and electrode interface effects in the frequency domain. Generally, the data in the complex plane is represented in any of the four basic formalisms. These are complex impedance ($Z^*$), complex admittance ($Y^*$), complex permittivity ($\varepsilon^*$), complex electric modulus ($M^*$), which are related to each other:

$Z^* = Z' - jZ'' = 1/j\omega C_0 \varepsilon^*$ (1)

$M^* = M' + jM'' = 1/\varepsilon^* = j\omega C_0 Z^*$ (2)

$Y^* = Y' + jY'' = j\omega C_0 \varepsilon^*$ (3)

$\varepsilon^* = \varepsilon' - j\varepsilon''$, $\tan\delta = \varepsilon''/\varepsilon' = M''/M' = Z'/Z'' = Y'/Y''$ (4)

Where ($Z'$, $M'$, $Y'$, $\varepsilon'$) and ($Z''$, $M''$, $Y''$ $\varepsilon''$) are real and imaginary components of impedance, electrical modulus, admittance and permittivity respectively, $\omega = 2\pi f$ is the angular frequency and $j = \sqrt{-1}$ the imaginary factor. The complex impedance of "electrode/sample/electrode" configuration can be explained as the sum of a single with a parallel combination of RC (R=resistance, C=capacitance) circuit. Thus, the result obtained using impedance analysis is an unambiguous, and hence provide a true picture of the electrical behavior of the material.

Figure 2 (a-d) shows the complex impedance spectrum ($Z'$ Vs. $Z''$) of $BN_xF_{1-x}$-PT with x=0.05, 0.10, 0.15, 0.20 at different temperatures. Single semicircular arcs exist in a wide temperature (200-275°C) region for different compositions. This confirms the presence of grain effect in the materials even if increasing the percentage of Nd concentration. It is also observed that as the temperature increases the intercept point on the real axis shifts towards the origin which indicates that decrease in the resistive property i.e., called bulk resistance ($R_b$) of the materials [22, 23]. The electrical process taking place within the material can be modeled (as an

RC circuit) on the basis of the brick-layer model [21]. The impedance data did not fit well with single RC-combination, rather this fit excellently well with equivalent circuits (insets of Fig. 2(a-d)) at $200^0C$ for x=0.05-0.20, and the fitting parameters of the circuit are shown in the table 1.

Figure 3 (a-d) shows the variation of Z′ as a function of frequency ($10^2$-$10^6$ Hz) of $BN_xF_{1-x}$-PT with x=0.05, 0.10, 0.15, 0.20 at different temperatures. It is observed that the magnitude of Z' (bulk resistance) decreases on increasing temperature as well as Nd concentration in the low frequency ranges (up to a certain frequency), and thereafter appears to merge in the high-frequency region. This is possible due to the release of space charge polarization with rise in temperatures and frequencies [24]. This behavior shows that the conduction mechanism increases with increasing temperature and frequency (i.e., negative temperature coefficient of behavior like that of a semiconductor). The coincidence of the value of Z' at higher frequencies at all the temperatures indicates a possible release of space charge [25, 26] and the frequency at which the release of space charge occurs also depends upon the Nd concentration. The space charge polarization occurs maximum at higher frequency side for x=0.20 concentration as compared to all other concentration. This may be due to the reduction in barrier properties of the materials with rise in temperature which is responsible for the enhancement of conductivity of the materials [27, 28]. At a particular frequency, Z' becomes independent of frequency. This type behavior is similar to the other material reported by Singh et. al. [29].

Figure 4 (a-d) shows the frequency-temperature dependence of Z" (usually called as loss spectrum) of $BN_xF_{1-x}$-PT with x=0.05-0.20. The magnitude of Z" decreases with increase in frequency as well as Nd concentration at high temperature region. The appearance of peaks in the loss spectrum at the high temperature region suggests the existence of relaxation process of the different compositions. This may due to the immobile species at low temperatures and defect or vacancies at high temperatures [29, 30]. It shows that with increasing Nd concentration and temperature, the magnitude of Z" decreases and all the peaks shift towards the higher frequency side and finally, all the curves merge in high-frequency region. At higher frequencies, the contribution from the grain predominates owing to the absence of the space charge effects of the different compositions [31]. Generally, the impedance data were used to evaluate the relaxation time (τ) of the electrical phenomena in the different compositions using the relation τ =

$1/\omega=1/2\pi f_r$, $f_r$ is the relaxation frequency. The variation of $\tau$ as a function of inverse of absolute temperature shown in the figure 5 calculated from impedance plot and it appears to be linear which follows the Arrhenius relation $\tau = \tau_0 \exp(-E_a/K_B T)$, where $\tau_0$ is the pre-exponential factor, $E_a$ is the activation energy, $K_B$ is the Boltzmann constant and T is the absolute temperature. The activation energy calculated from the slope of log ($\tau$) against $10^3/T$ ($K^{-1}$) curves of $BN_xF_{1-x}$-PT are 0.89, 0.76, 0.71, 0.70 eV for x=0.05, 0.10, 0.15 and 0.20 respectively.

## 4 Modulus Properties

Figure 6 (a-d) shows the complex modulus spectrum (M' Vs M'') of $BN_xF_{1-x}$-PT with x=0.05, 0.10, 0.15, 0.20 at different temperatures. Complex impedance spectrum gives more emphasis to elements with larger resistance whereas complex electric modulus plots highlight those with smaller capacitance. Using the complex electric modulus formalism, the inhomogeneous nature of the polycrystalline sample can be probed into bulk and grain boundary effects, which may not be distinguished from complex impedance plots. The complex electric modulus spectrum of the samples does not form semicircles. But slightly they exhibit deformed arcs with their centers positioned below the x-axis. The scaling behavior in the sample was studied by plotting $M''/M''_{max}$ versus $\log_{10} f/f_{max}$ at different temperatures for different compositions as shown in the figure 6(a-d) (inset). This is called the master modulus curve. It gives an insight into the dielectric processes occurring inside the material. It indicates a slight shift in the peak pattern and similar shape of spectrum with a slight variation in the full-width at half maximum (FWHM) with rise in temperature.

Figure 7(a-d) shows the variation of M' as a function of frequency for $BN_xF_{1-x}$-PT with x=0.05, 0.10, 0.15, 0.20) at selected temperatures. For all the concentrations Nd, show that M' approaches to zero in the low frequency region, and a continuous dispersion on increasing frequency may be contributed to the conduction phenomena due to short range mobility of charge carriers. This implies the lack of a restoring force for flow of charge under the influence of a steady electric field [32]. This confirms elimination of electrode effect in the material.

Figure 8(a-d) shows the variation of imaginary part of electric modulus with frequency for $BN_xF_{1-x}$-PT (x=0.05-0.20) at selected temperatures. The maxima $M''_{max}$ shifts towards higher frequencies side with rise in temperature as well as Nd concentration ascribing correlation

between motions of mobile ions [33] and suggests that the dielectric relaxation is thermally activated process. The asymmetric peak broadening indicates the spread of relaxation times with different time constant, and hence relaxation is of non- Debye type. The low frequency peaks show that the ions can move over long distances whereas high-frequency peaks merge to spatially confinement of ions in their potential well. The nature of modulus spectrum suggests the existence of hopping mechanism of electrical conduction in the materials. The variation of relaxation time ($\tau$) as a function of reciprocal of temperature 1/T of $BN_xF_{1-x}$-PT (x=0.05-0.20) at high temperature region is shown in Fig. 9 (calculated from modulus spectrum). This graph follows the Arrhenius relation, $\tau = \tau_0 \exp(-E_a/K_BT)$ where the symbols have their usual meanings and thermally activated process. The activation energy of the composites is found to be 0.81, 0.69, 0.68 and 0.67eV for different compositions.

## 5 Electrical conductivity

The ac electrical conductivity ($\sigma_{ac}$) was calculated from the dielectric data using the empirical relation: $\sigma_{ac}=\omega\varepsilon_r\varepsilon_0\tan\delta$, where $\varepsilon_0$ is the vacuum dielectric constant and $\omega$ (=$2\pi f$) is the angular frequency. Figure 10 shows the variation of $\sigma_{ac}$ with inverse of absolute temperature ($10^3$/T) of $BN_xF_{1-x}$-PT (x=0.05-0.20) at a particular frequency (1MHz). The nature of variation of $\sigma_{ac}$ over a wide temperature range supports the thermally activated transport properties of the materials obeying Arrhenius equation $\sigma_{ac}=\sigma_0\exp(-E_a/KT)$, where $\sigma_0$ is the pre-exponential factor and other symbols have their usual meaning. It is observed that the ac conductivity of the material increases with rise in temperature and show the NTCR behavior. This behavior suggests that the conduction mechanism of the compounds may be due to the hopping of charge carrier. The values of activation energy of the composites are found to be 0.28, 0.27, 0.31and 0.32 eV for different compositions x=0.05-0.20 respectively.

The frequency–temperature dependence of ac conductivity is shown in Figure 11(a-d). This figure shows that the conductivity for the concentration x=0.10, 0.15, 0.20 exhibit step like decrease in low frequency region suggest low frequency polarization which may be due to the transition from the bulk to contact resistance [25]. There is a deviation in the slope at a particular frequency usually called hopping frequency. This typical behavior suggests the presence of hopping mechanism in the composite. This type of conducting behavior is well described by

Jonscher's universal power law [34]: $\sigma(\omega) = \sigma_{dc} + A\omega^n$ where $n$ is the frequency exponent with $0 < n < 1$ and A is the temperature dependent pre-exponential factor. The term $A\omega^n$ comprises the ac dependence and characterizes all dispersion phenomena. The exponent $n$ can vary differently from material to material, depending on temperature. This suggests that the electrical conduction in $BN_xF_{1-x}$-PT is a thermally activated process. The non linear curve well fit to Jonscher's power law for all the composites with x=0.05, 0.10, 0.15, 0.20 at different temperature (200-275) shown in the figure 11(a-d). The fitting parameter A, n are calculated from the non-linear fitting are given in the table 2.

Figure 12 shows the variation of $\sigma_{dc}$ (bulk) with inverse of absolute temperature ($10^3/T$) of $BN_xF_{1-x}$-PT (x=0.05-0.20). The dc conductivity ($\sigma$) of the material was evaluated by using the relation $\sigma = l/R_bA$, where l is the thickness, $R_b$ is the bulk resistance (calculated from impedance plot) and A is the area of cross section of the sample. The dc conductivity is found to be increase with increase in Nd concentration. The nature of the plot may be explained by Arrhenius type relation: $\sigma_{dc} = \sigma_0 \exp(-E_a/KT)$, where $\sigma_0$ is the pre- exponential factor. A linear fit of $\sigma_{dc}$ versus ($10^3/T$) plot has been used to evaluate the activation energy $E_a$ and found as 0.73, 0.76, 0.71, 0.73eV for x=0.05, 0.10, 0.15, 0.20 respectively in the temperature region (200-275$^0$C).

## 6 Conclusions

The $BN_xF_{1-x}$-PT (x = 0.05, 0.10, 0.15, 0.20) composites were prepared by a high-temperature solid state reaction technique. X-ray structural analysis confirmed the materials show tetragonal structure at room temperature. Complex impedance spectroscopy was used to characterize the electrical properties of the materials. The electrical conduction in the composites is due to the bulk effects only. The bulk resistance decreases with rise in temperature and exhibit a NTCR behavior. The electrical modulus has confirmed the presence of a hopping mechanism in the materials. The dc conductivity shows a typical Arrhenius type of electrical conductivity. The activation energy calculated both from impedance and modulus spectra are comparable and is of the same order. The ac conductivity spectrum was found to obey Jonscher's universal power law.

**Acknowledgement**

The authors acknowledge the financial support through DRS-I of UGC under SAP for the development of research work in the School of Physics, Sambalpur University and to the DST under SERC Fast Track Scheme for Young Scientist (Project No. SR/FTP/PS-036/2011) New Delhi, India.

**References**

[1] G.A. Smolenskii, and I.E. Chupis, Sov. Phys. Uspekhi **25** (7), 475 (1982).

[2] Yu.N. Venevtsev, and V.V. Gagulin, Ferroelectrics **162,** 23-31 (1994).

[3] N.A. Hill, J. Phys. Chem. B **104** (29), 6694–6709 (2000).

[4] N.A. Spaldin, and M. Fiebig, Science **309** (5733), 391–392 (2005).

[5] J. Wang, J.B. Neaton, H. Zheng, V. Nagarajan, S.B. Ogale, B. Liu, D. Viehland, V. Vaithyanathan, D.G. Schlom, U.V. Waghmare, N.A. Spaldin, K.M. Rabe, M. Wuttig, and R. Ramesh, Science **299** (5613), 1719-1722 (2003).

[6] N.A. Hill, and A. Filippetti, J. Magn. Magn. Mater **242–245,** 976–979 (2002).

[7] M. Fiebig, J Phys D: Appl Phys **38,** R123–R152 (2005).

[8] R. Ramesh, and N.A. Spaldin, Nature Materials **6**, 21-29 (2007); S. W. Cheong, and M. Mostovory, Nature Materials **6**, 13-20 (2007); W. Eerenstein, N. D. Mathur, and J.F. Scott, Nature (London) **442,** 759-765 (2006).

[9] J. Silva, A. Reyes, H. Esparza, H. Camacho, and L. Fuentes, Integrated Ferroelectrics **126,** 47-59 (2011).

[10] S.T. Zhang, Y. Zhang, M.H. Lu, C.L. Du, Y.F. Chen, Z.G. Liu, Y.Y. Zhu, and N.B. Ming, Appl. Phys. Lett. **88,** 162901 (2006).

[11] Y. Zhang, S. Yu, and J. Cheng, J. Euro. Ceramic Soc. **30,** 271 (2010).

[12] C.F. Chung, J.P. Lin, and J.M. Wu, Appl. Phys. Lett. **88,** 242909 (2006).

[13] X. Qi, J. Dho, R. Tomov, M.G. Blamire, and J.L. MacManus-Driscoll, Appl. Phys. Lett. **86**, 062903 (2005).

[14] M.K. Singh, H.M. Jang, S. Ryu, and M.H. Jo, Appl. Phys. Lett. **88,** 042907 (2006).

[15] G.L. Yuan, S.W. Or, and H.L.W. Chan, J. Appl. Phys. **101,** 064101 (2007).

[16] Yu.N. Venevstev, G.S. Zhdanov, S.N. Solov'ev, E.V. Bezus, V.V. Ivanova, S.A. Fedulov, and A.G. Kapyshev, Sov. Phys. Crystallogr. **5,** 594 (1960).

[17] N. Wang, J. Cheng, A. Pyatakov, A.K. Zvezdin, J.F. Li, L.E. Cross, and D. Viehland, Phys. Rev. B **72,** 104434 (2005).

[18] W. M. Zhu, H. Y. Guo, and Z. Ye, Phys. Rev. B **78,** 014401 (2008).

[19] V.V.S.S. Sai Sunder, J. Mater. Res **10(5),** 1301 (1995).

[20] Wu. E., J. Appl. Cryst. **22**, 506 (1989).

[21] J.R. Mac Donald, Impedance Spectroscopy, Wiley (New York) (1987).

[22] B. Behera, P. Nayak, and R.N.P. Choudhary, J. Mater Sci: Mater Electron **19,** 1005-1011 (2008).

[23] D.C. Sinclair, and A.R. West, J. Appl. Phys. **66,** 3850-3856 (1989).

[24] R.N.P. Choudhary, D.K. Pradhan, C.M. Tirado, G.E. Bonilla, and R.S. Katiyar, Physica B **393**, (1-2), 24–31 (2007).

[25] B. Behera, P. Nayak, and R.N.P. Choudhary, Cent. Eur. J. Phys. **6(2),** 289-295 (2008).

[26] J. Plocharski, and W. Wieczoreck, Solid State Ion **28-30,** 979-982 (1988).

[27] V. Provenzano, L.P. Boesch, V. Volterra, C.T. Moynihan, and P.B. Macedo, J. Am. Ceram. Soc. **55,** 492-496 (1972).

[28] H. Jain, and C.H. Hsieh, J. Non-Cryst. Solids **172-174,** 1408-1412 (1994).

[29] H. Singh, A. Kumar, and K.L. Yadav, Materials Science and Engineering B **176,** 540-547 (2011)

[30] B. Behera, P. Nayak, and R.N.P. Choudhary, Materials Research Bulletin **43,** 401- 410 (2008).

[31] B. Behera, P. Nayak, and R.N.P. Choudhary, Materials Chemistry and Physics **106,** 193-197 (2007)

[32] P.B. Macedo, C.T. Moynihan, and R. Bose, Phys. Chem. Glasses **13,** 171 (1972).

[33] F. Borsa, D.R. Torgeson, S.W. Martin, and H.K. Patel, Phy. Rev. B **46,** 795 (1992).

[34] A. K. Jonscher, Nature **267,** 673-679 (1977).

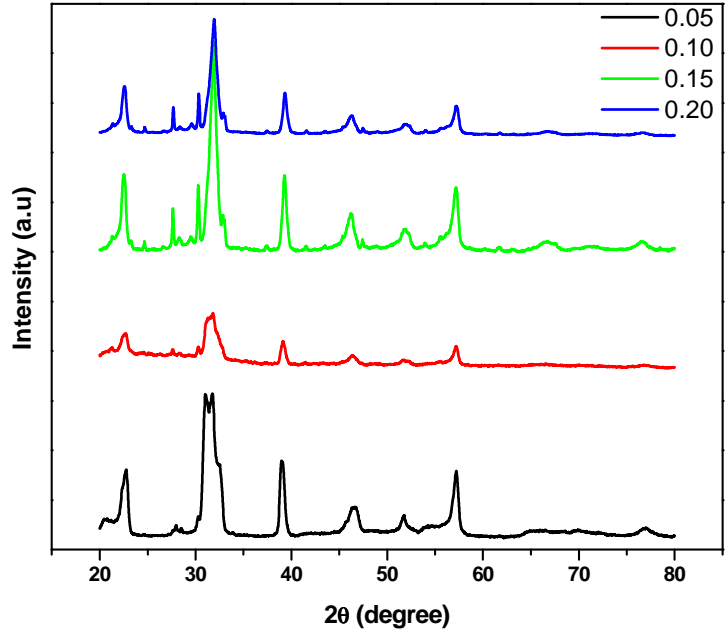

**Figure 1** X-Ray diffraction patterns of 0.5BiNd$_x$Fe$_{1-x}$O$_3$-0.5PbTiO$_3$ (BN$_x$F$_{1-x}$-PT) (x=0.05-20) at room temperature.

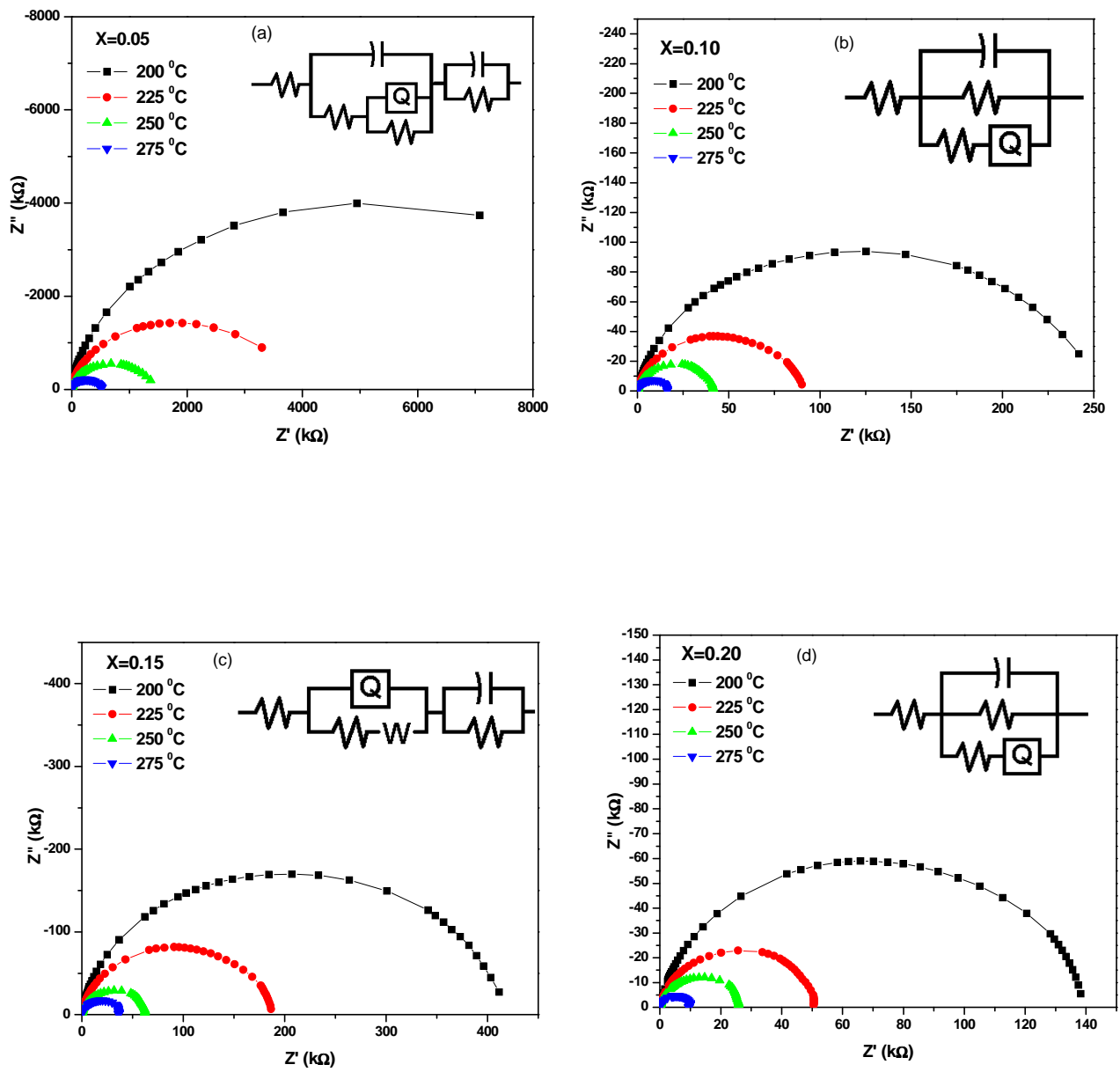

**Figure 2 (a-d)** Complex impedance plot (Z' Vs Z'') in the temperature range 200-275$^{o}$C.

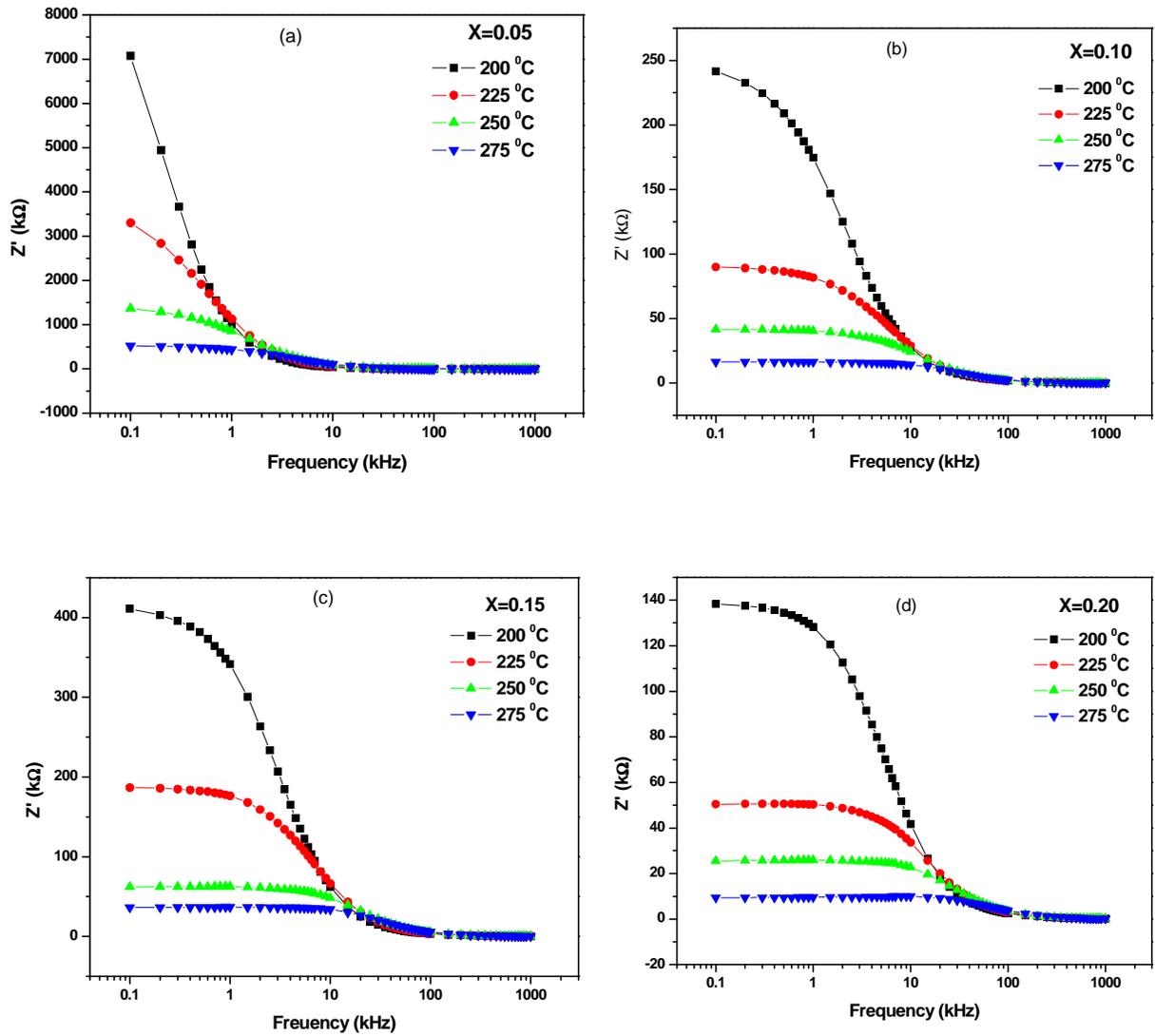

**Figure 3 (a-d)** Variation of Z' with frequency at various temperatures.

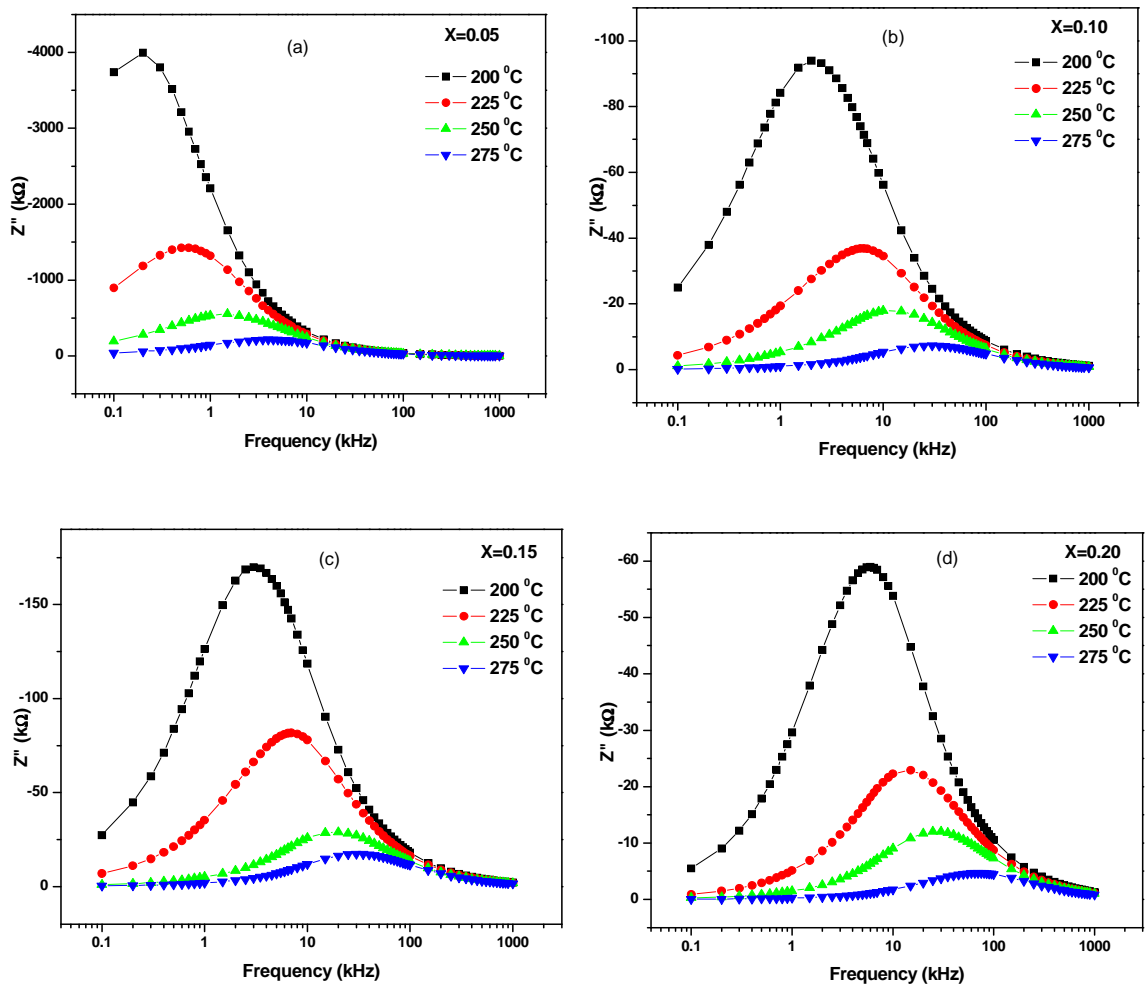

**Figure 4 (a-d)** Variation of Z" with frequency at various temperatures.

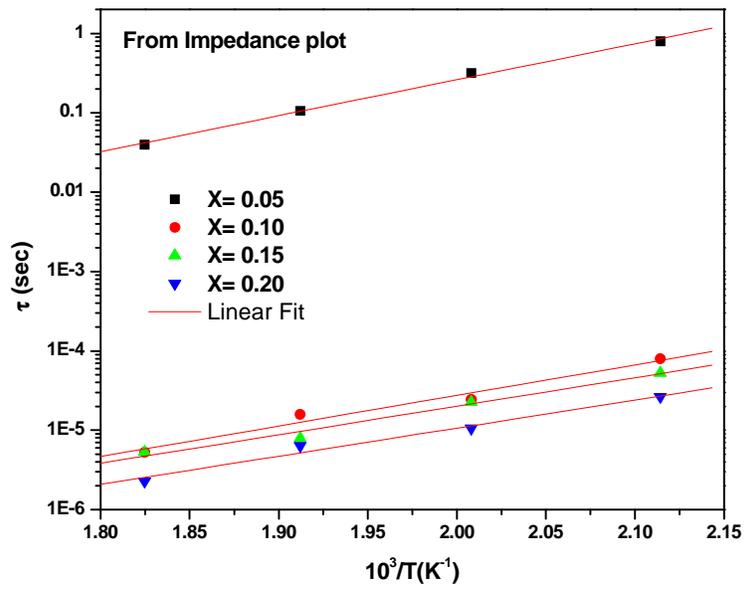

**Figure 5** Variation of relaxation time with inverse of temperature ($10^3/T$) calculated from impedance plot.

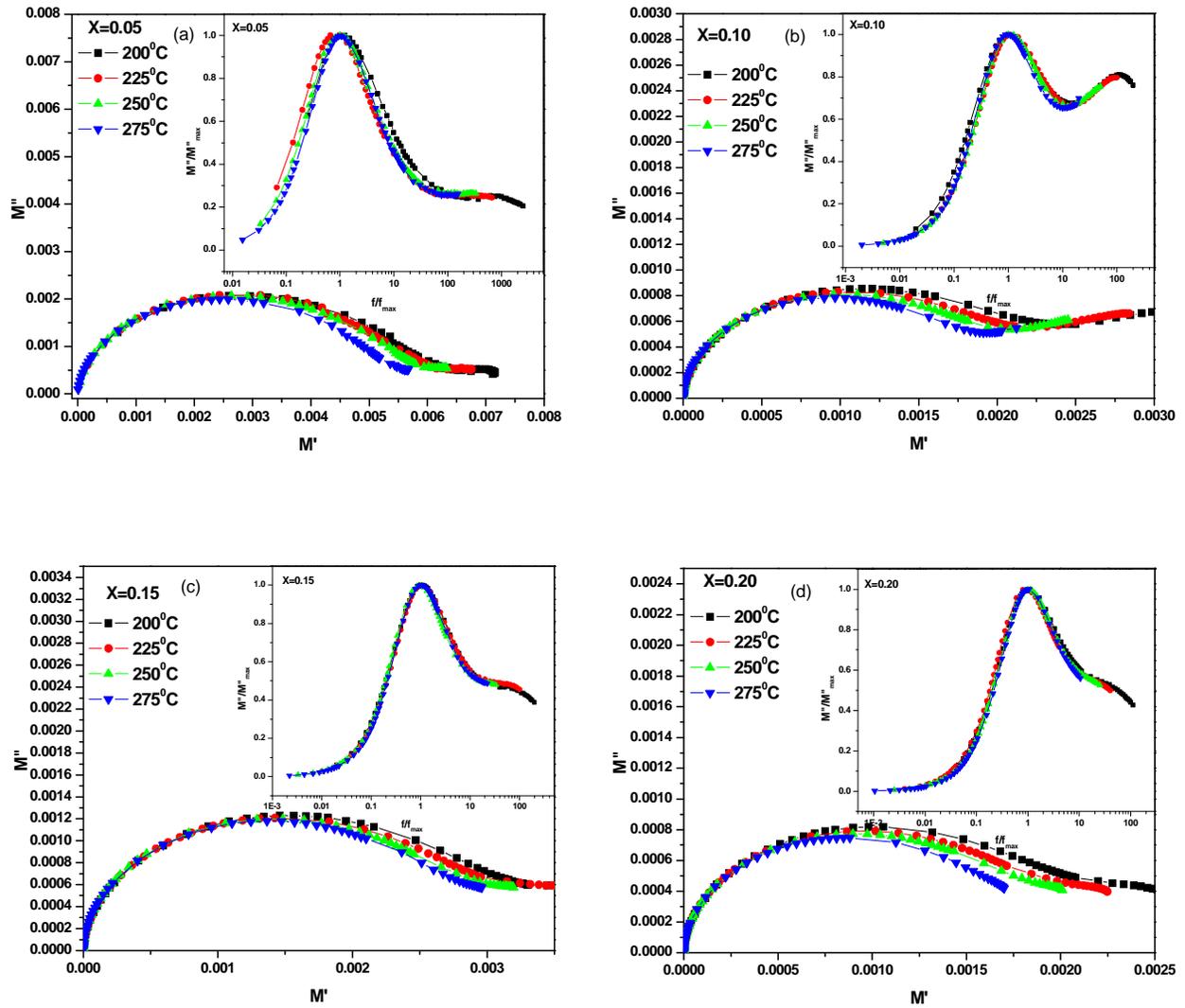

**Figure 6 (a-d)** Complex modulus spectrum (M' Vs M") and the variation of M"/M"$_{max}$ versus $\log_{10}$ f/f$_{max}$ (inset) of BN$_x$F$_{1-x}$-PT with x=0.05-0.20 at different temperatures.

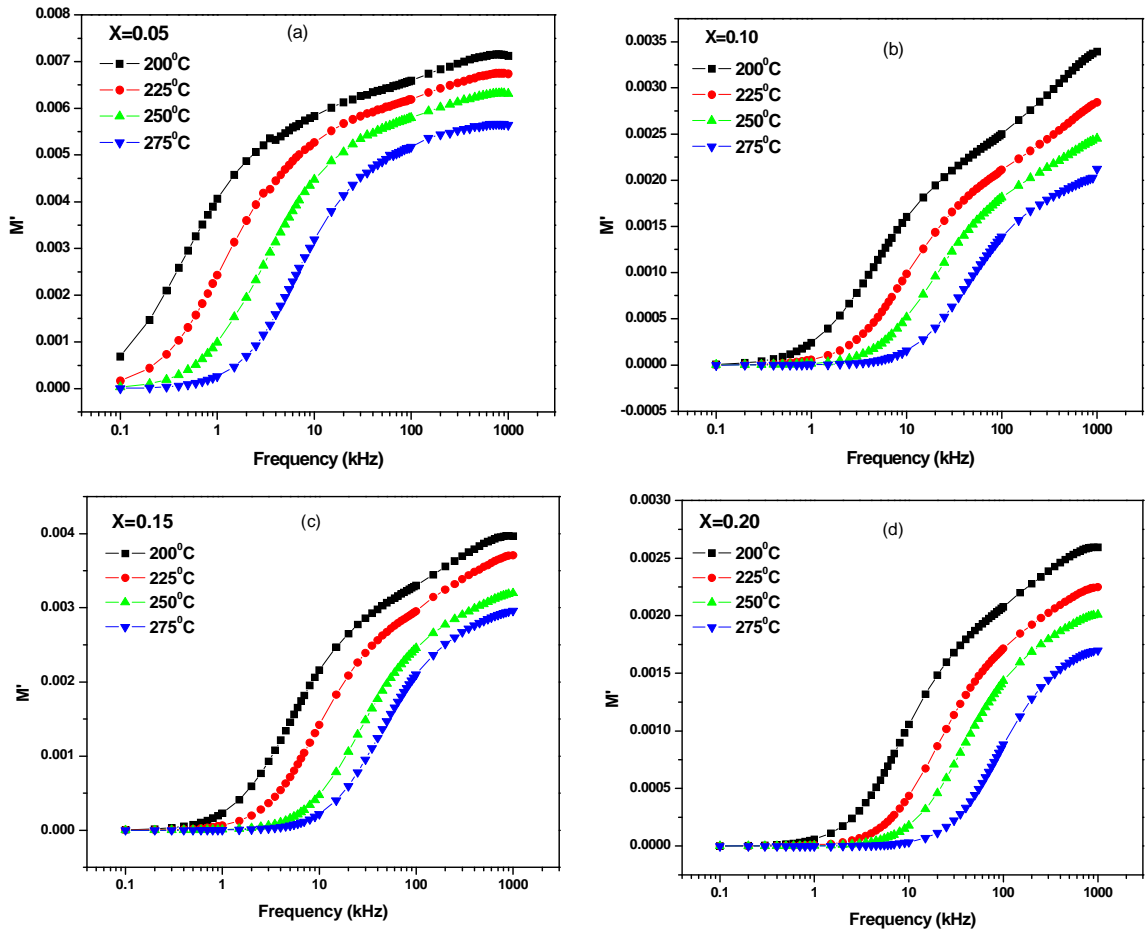

**Figure 7 (a-d)** Variation of M' as a function of frequency for $BN_xF_{1-x}$-PT with x=0.05-0.20 at selected temperatures.

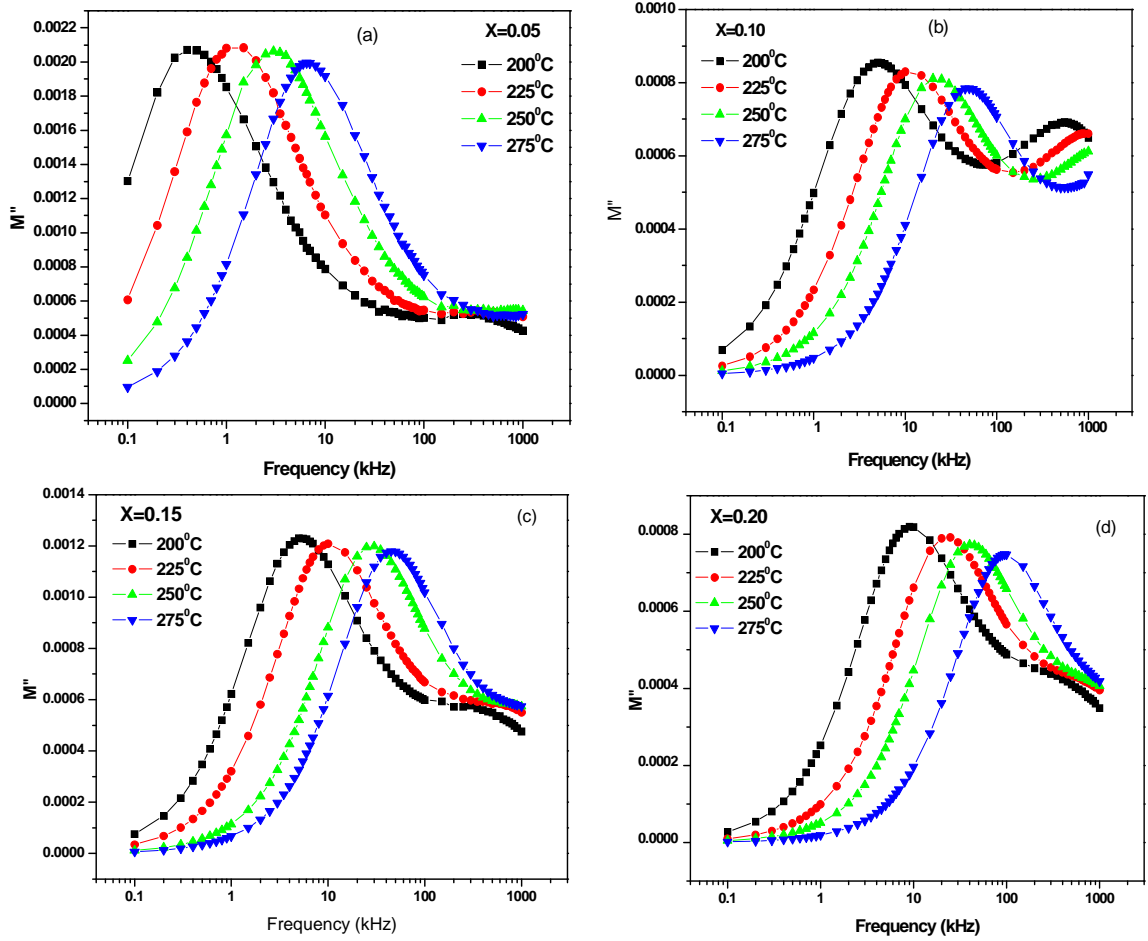

**Figure 8 (a-d)** Variation of imaginary part of electric modulus with frequency for $BN_xF_{1-x}$-PT (x=0.05-0.20) at selected temperatures.

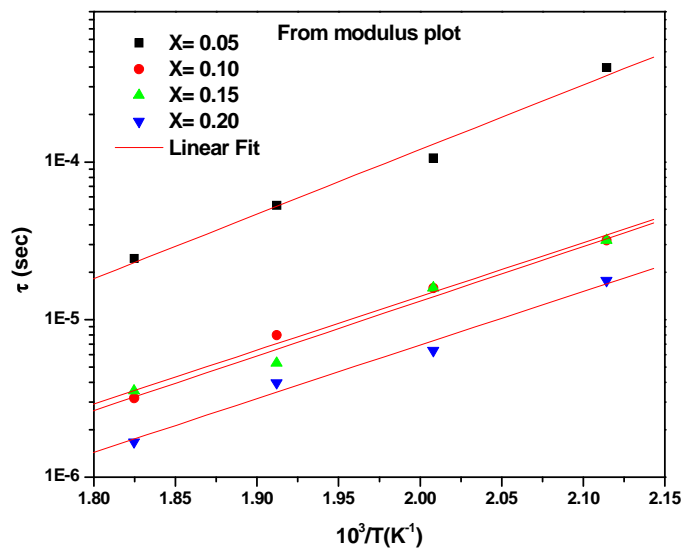

**Figure 9** Variation of relaxation time (τ) as a function of reciprocal of temperature 1/T of $BN_xF_{1-x}$-PT (x=0.05-0.20) at different temperature region calculated from modulus spectrum.

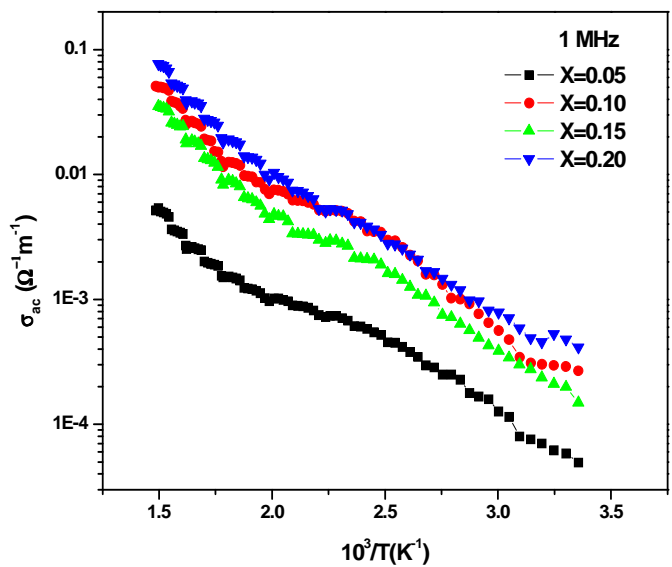

**Figure 10** Variation of ac conductivity with inverse of temperature ($10^3$/T).

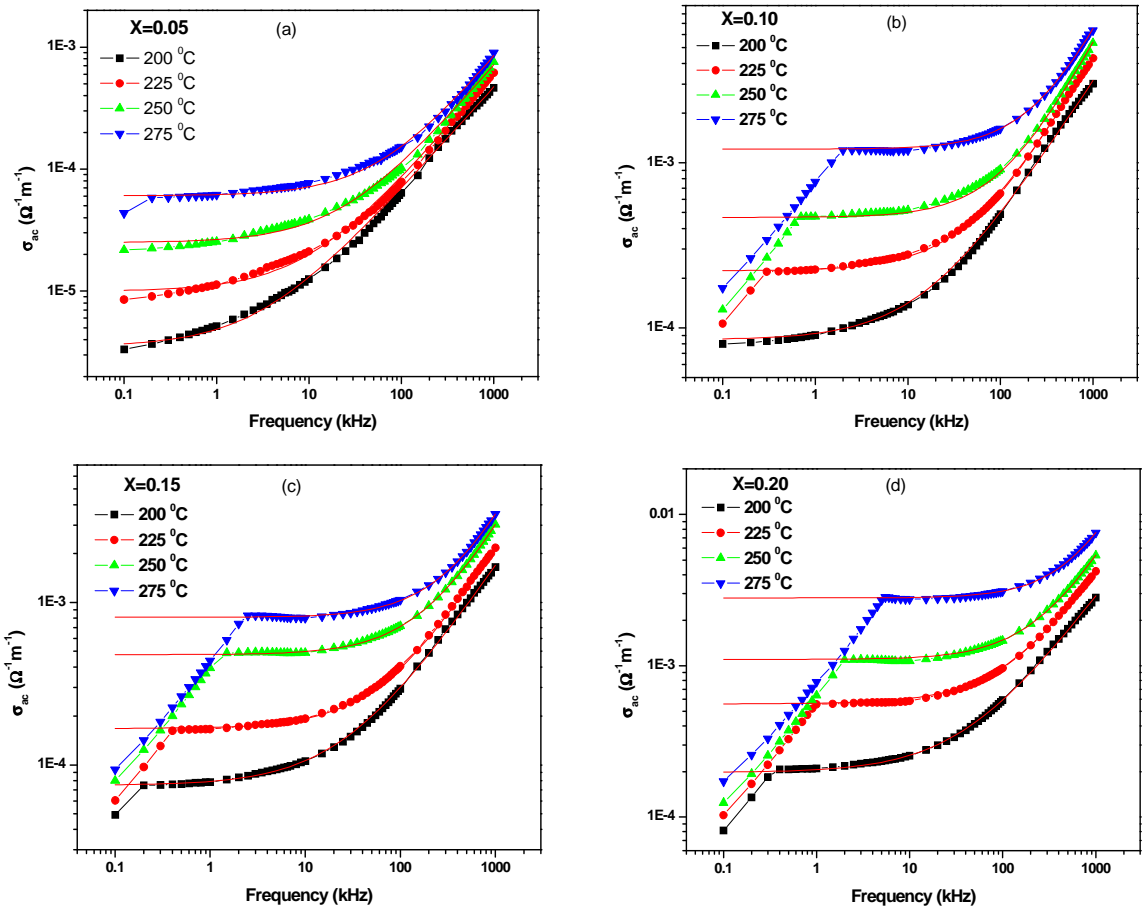

**Figure 11(a-d)** Variation of ac conductivity with frequency at different temperature

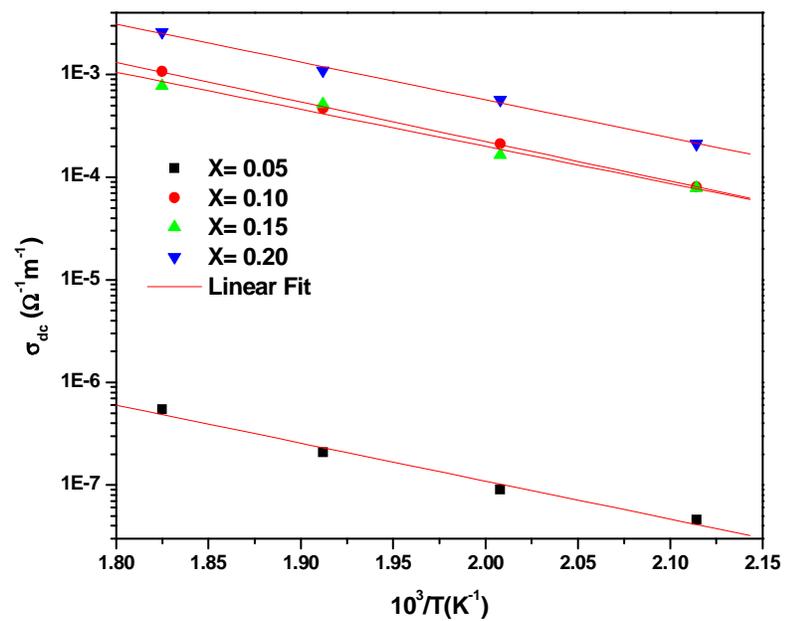

**Figure 12** Variation of dc conductivity with inverse of temperature.

| Different Parameters | X=0.05 | X=0.10 | X=0.15 | X=0.20 |
|---|---|---|---|---|
| $R_1$ | 16517 | 9.9E-8 | 1.001E-7 | 51.67 |
| $R_2$ | 1.225E-5 | 4.546E-4 | 2.83E-5 | 1.434E-5 |
| $R_3$ | 1.172E-7 | 2.06E-5 | 1.176E-5 | 8588 |
| $R_4$ | 1476 | | | |
| $C_1$ | 4.364E-11 | 7.028E-9 | 8.211E-10 | 1.116E-10 |
| $C_2$ | 5.52E-10 | | | |
| CPE | 1.155E-9 | 9.885E-10 | 3.228E-10 | 1.802E-9 |
| Frequency power (n) | 0.5493 | 0.8718 | 0.9089 | 0.7211 |
| Chi square | 0.002373 | 0.0156 | 0.01227 | 0.01415 |
| Warborg ($\gamma_0$) | | | 1.413E-6 | |

Table 1 Summarizing of fitting parameters corresponding to equivalent circuits at $200^0C$ of Fig. 2 (a-d).

| T($^o$C) | x=0.05 | | x=0.10 | | x=0.15 | | x=0.20 | |
|---|---|---|---|---|---|---|---|---|
| | A | n | A | n | A | n | A | n |
| 200 | $3.94\times10^{-9}$ | 0.84 | $1.93\times10^{-8}$ | 0.86 | $1.17\times10^{-8}$ | 0.85 | $2.75\times10^{-8}$ | 0.83 |
| 225 | $2.80\times10^{-9}$ | 0.88 | $6.5\times10^{-9}$ | 0.96 | $3.64\times10^{-9}$ | 0.95 | $8.00\times10^{-9}$ | 0.94 |
| 250 | $3.18\times10^{-9}$ | 0.88 | $1.71\times10^{-9}$ | 1.07 | $1.54\times10^{-9}$ | 1.03 | $1.98\times10^{-9}$ | 1.05 |
| 275 | $2.16\times10^{-9}$ | 0.92 | $1.00\times10^{-9}$ | 1.12 | $7.11\times10^{-10}$ | 1.09 | $3.42\times10^{-10}$ | 1.19 |

Table 2 Fitting parameters obtained from the Jonscher's power law at different temperatures.